\documentclass[conference]{IEEEtran}
\IEEEoverridecommandlockouts
\usepackage{cite}
\usepackage{amsmath,amssymb,amsfonts}
\usepackage{algorithmic}
\usepackage{graphicx}
\usepackage{textcomp}
\usepackage{xcolor}
\usepackage{booktabs}
\def\BibTeX{{\rm B\kern-.05em{\sc i\kern-.025em b}\kern-.08em
    T\kern-.1667em\lower.7ex\hbox{E}\kern-.125emX}}
\begin{document}

\title{Knowing the Rules Is Not Enough: \\Student Regulatory Awareness and Use of GenAI in Higher Education}

\author{

\begin{minipage}[t]{0.48\textwidth}
\centering
1\textsuperscript{st} Lasse Bischof\\
\textit{Dpt. of Business Information Systems}\\
\textit{University of Applied Sciences and Arts Hannover}\\
Hannover, Germany\\
0009-0002-6622-0770
\end{minipage}
\hfill
\begin{minipage}[t]{0.48\textwidth}
\centering
2\textsuperscript{nd} Eva-Maria Schön\\
\textit{University of Applied Sciences Emden/Leer}\\
Emden/Leer, Germany\\
0000-0002-0410-9308
\end{minipage}

\vspace{0.6cm}

\and

\begin{minipage}[t]{0.48\textwidth}
\centering
3\textsuperscript{rd} Maria Rauschenberger\\
\textit{University of Applied Sciences Emden/Leer}\\
Emden/Leer, Germany\\
0000-0001-5722-576X
\end{minipage}
\hfill
\begin{minipage}[t]{0.48\textwidth}
\centering
4\textsuperscript{th} Michael Neumann\\
\textit{Dpt. of Business Information Systems}\\
\textit{University of Applied Sciences and Arts Hannover}\\
Hannover, Germany\\
0000-0002-4220-9641
\end{minipage}
}





\maketitle

\begin{abstract}
\textit{Context:} Generative Artificial Intelligence (GenAI) tools such as ChatGPT are increasingly integrated into students’ learning practices. While previous research mainly examines adoption rates and attitudes, students’ awareness of institutional regulations and their perceived compliance remain unexplored. Understanding whether regulatory awareness influences student behavior is therefore important as higher education institutions create and apply AI policies. 
\textit{Objective:} This study investigates how students’ awareness of GenAI regulations relates to their perceived compliance and actual usage behavior. Our research objective is to examine the association between regulatory knowledge, GenAI use, and perceived rule conformity among students in computer science–related study programs. 
\textit{Method:} A survey with 151 undergraduate students in Business Information Systems and E-Government programs at the University of Applied Sciences and Arts Hannover (Germany) collected data on GenAI usage, tools used, awareness of institutional regulations, and perceived compliance. Descriptive statistics, cross-tabulations, and correlation analyzes were applied. 
\textit{Results:} Most students actively use GenAI tools, but over half are uncertain whether their usage complies with institutional regulations. Regulatory awareness shows only weak to moderate associations with actual usage behavior. Students primarily rely on privately accessed GenAI tools rather than institutionally provided solutions. 
\textit{Contributions:} The study contributes empirical evidence on the relationship between regulatory awareness and GenAI usage in higher education. Our findings highlight a gap between institutional regulations and student practices and provide insights for educators and institutions on improving policy communication and integrating GenAI more effectively into teaching and learning contexts. 
\end{abstract}


\section{Introduction}
The rapid emergence of generative artificial intelligence (GenAI) tools such as ChatGPT have begun to reshape how students learn, access information, and complete academic tasks in higher education~\cite{Schoen.2023}. These tools are increasingly integrated into students’ everyday learning practices, particularly in computer science–related study programs, where they are frequently used for tasks such as coding support, problem solving, and text generation (\textit{e.g.}, \cite{Bischof.2025,Wang.2025}).

The widespread availability of GenAI tools has triggered an ongoing debate in higher education regarding both their educational potential and the risks they may introduce~\cite{Mah.2024}. On the one hand, these technologies may support learning processes by providing explanations~\cite{Daun.2023}, assisting with programming tasks (\textit{e.g.}, \cite{Maher.2023,Speth.2023}), or enabling new forms of knowledge exploration~\cite{Lauren.2023}. On the other hand, educators have raised concerns about academic integrity, authorship, and the potential misuse of AI-generated content in assessments (\textit{e.g.}, \cite{Aldossary.2024,neumann_we_2023}). As a consequence, many universities have begun developing institutional regulations and guidelines intended to govern the acceptable use of GenAI tools in academic contexts (\textit{e.g.}, \cite{Rizki.2025,Shailendra.2024,Southworth.2023}). Also, the importance of ethical considerations and the awareness of academic integrity arises~\cite{Wiese.2025}.

However, the existence of institutional policies does not necessarily imply that students are aware of these regulations or that such policies meaningfully influence their behavior. In practice, regulatory guidelines are often communicated inconsistently across courses or remain ambiguous from the students’ perspective~\cite{Su.2023}. This may lead to uncertainty about whether and how GenAI tools may be used in academic work. As a result, students may either avoid potentially useful technologies or use them in ways that are not aligned with institutional expectations.

While prior research has primarily examined adoption rates, attitudes toward GenAI~\cite{Li.2025}, and common use cases among students~\cite{DeFrancisis.2025}, considerably less attention has been paid to how students perceive and interpret institutional regulations governing the use of these technologies. Despite the growing number of institutional AI policies in higher education, empirical insights into how these regulations are perceived by students remain limited. In particular, little empirical evidence exists on whether students are aware of institutional GenAI policies and whether such awareness influences their usage behavior in practice.

This gap is particularly relevant for software engineering~\cite{Groepler.2025} and computer science education, where GenAI tools are frequently used for tasks such as programming support~\cite{Speth.2023}, problem solving~\cite{Tarun.2025}, and technical writing~\cite{Mendonca.2025}. As universities increasingly introduce policies to regulate AI-assisted learning, understanding whether these regulations are actually perceived and interpreted by students becomes critical for designing effective governance strategies for AI-supported learning environments.

To address this gap, this paper investigates the following research questions:
\begin{itemize}
    \item \textbf{RQ1:} To what extent are students aware of institutional regulations governing the use of GenAI tools in higher education?\\
    The first research question addresses the extent to which students are aware of institutional regulations governing the use of GenAI tools in higher education. It focuses on whether such regulations are actually perceived by students in their study context.
    \item \textbf{RQ2:} How do students perceive the rule-compliance of their own use of GenAI tools in their academic work?\\
    This research question examines how students assess the rule-compliance of their own use of GenAI tools in academic work. It therefore captures students’ self-perceived ability to judge whether their usage aligns with institutional expectations.
    \item \textbf{RQ3:} What relationship exists between students’ awareness of institutional GenAI regulations and their perceived rule-compliant use of these tools?\\
    The third research question explores the relationship between students’ awareness of institutional GenAI regulations and their perceived rule-compliant use of these tools. In doing so, it examines whether awareness of institutional rules is reflected in greater clarity regarding acceptable use in practice.
\end{itemize}

This study provides empirical insights into the gap between institutional regulations and students’ learning practices by examining the relationship between regulatory awareness and actual AI usage. The findings aim to inform educators and higher education institutions about how regulatory frameworks (such as guidelines for thesis or rules for the use of tools) are perceived by students and to support the development and improvement of clear and effective approaches for integrating GenAI tools into teaching and learning environments.

This paper is structured as follows: In Section~\ref{sec2:RelatedWork}, we give a brief overview of the work closely related to the topic of our paper. Next, in Section~\ref{sec3:ResearchDesign} we explain the research design including the applied data collection and analysis approach. Based on the presented results, we answer our research questions in Section~\ref{sec4:Results} and discuss the practical implications in Section~\ref{sec5:Discussion}. The paper closes with the conclusion and an outlook of our future work activities in Section~\ref{sec6:Conclusion}. 

\section{Related Work}
\label{sec2:RelatedWork}
Research on GenAI in higher education has expanded rapidly in recent years~\cite{Chiu.2023,Wang.2024}. Existing studies have primarily examined students’ adoption of AI tools (\textit{e.g.},\cite{von_garrel_kunstliche_2023,gottschling_nutzung_2024,Speth.2023}), their attitudes toward such technologies (\textit{e.g.}, \cite{Bischof.2025,Ruedian.2025}), and the educational opportunities and risks associated with their use (\textit{e.g.}, \cite{Meissner.2024,Schoen.2023}). At the same time, an increasing number of contributions have begun to address students’ awareness of AI-related ethical, legal, and institutional issues. 

Several studies provide a useful starting point for understanding students’ awareness of AI-related issues more broadly. Ghotbi and Ho \cite{Ghotbi.2021}, for example, examine students’ moral awareness regarding artificial intelligence and show that ethical reflection is an important dimension of how students engage with AI technologies. Dergunova et al. \cite{Dergunova.2022} investigate students’ general awareness of artificial intelligence and report varying levels of understanding across the student population. Similarly, Simon et al. \cite{Simon.2024} analyze the role of generative AI awareness in shaping student perceptions and show that awareness affects not only familiarity with the technology itself but also students’ evaluation of its role in educational settings. Wu et al. \cite{Wu.2025} further extend this line of research by focusing on GenAI risk awareness among students and situating their findings within an AI literacy perspective.

Further studies have also addressed the ethical dimensions of AI from the perspective of students. Ferhataj et al. \cite{Ferhataj.2025} show that students associate AI development with broader societal and ethical concerns rather than treating it merely as a technical innovation. Similarly, Liu et al. \cite{Liu.2025} investigate students’ strategic use of artificial intelligence in self-regulated learning and emphasize that students’ interaction with AI tools is shaped by more than simple adoption or frequency of use. These contributions are valuable because they underline that awareness, perception, and reflective use are central aspects of AI in educational contexts. However, they do not specifically address how students perceive institutional regulations governing the use of GenAI tools in higher education.

At the same time, the regulatory perspective has started to attract greater attention. Miah et al. \cite{Miah.2025} discuss legal and ethical frameworks for deploying AI and machine learning in U.S. educational institutions and highlight the growing importance of institutional governance structures. Stephano \cite{Stephano_2025} approaches the issue from the perspective of lecturers’ copyright law awareness in regulating AI use by students. These studies show that regulation is becoming an important topic in higher education practice and policy. However, they mainly focus on legal and institutional frameworks or on the perspective of educators rather than on how such regulations are perceived by students themselves.

Among the studies most closely related to our work is Tan et al. \cite{Tan.2025}, who investigate student perceptions of GenAI regulations in higher education in Singapore. At the same time, their study also indicates that this line of research is still emerging. More generally, the existing literature provides valuable insights into AI awareness, ethical considerations, and regulatory developments, but it remains less clear whether students are actually aware of institutional GenAI regulations and whether such awareness is reflected in how they assess the rule compliance of their own usage behavior.

This gap is especially relevant in computer science related study programs, where GenAI tools are already used for various tasks such as programming support, problem solving, and text generation. 
It is particularly relevant in the context of European University contexts, where the use of GenAI in teaching and learning is shaped not only by institutional rules, but also by a dense external regulatory environment. 
In particular, the General Data Protection Regulation (GDPR) governs the processing of personal data across the EU, while the EU AI Act has introduced a directly applicable regulatory framework for AI systems, including obligations that are increasingly relevant for educational institutions and their use of AI-supported technologies. While prior work has helped to establish that awareness and regulation matter, there is still limited empirical evidence on the relationship between students’ awareness of institutional GenAI regulations and their perceived rule-compliant use of these tools. Our study addresses this gap by examining both dimensions together and by focusing on a higher education context in which GenAI tools are already embedded in students’ everyday academic practices and interpreted against a broader European regulatory backdrop.

\section{Research Design}
\label{sec3:ResearchDesign}
To investigate the relationship between students’ awareness of GenAI regulations and their usage behavior, we conducted a quantitative survey study. The research design focuses on examining students’ awareness of institutional regulations, their perceived rule-compliant use of GenAI tools, and the relationship between these aspects.

A survey-based approach was selected to obtain empirical insights into students’ current practices and perceptions regarding the use of GenAI tools in higher education. According to Kitchenham et al. \cite{Kitchenham.2022}, surveys are particularly suitable for capturing self-reported perceptions, attitudes, and behavioral patterns across a larger group of participants and therefore allow the systematic investigation of relationships between regulatory awareness and perceived rule-compliant behavior.

The following subsections describe the research context, the survey design, the data collection procedure, the applied analysis methods, and potential threats to validity.

\subsection{Research Context}
The study was conducted at the University of Applied Sciences and Arts Hannover in Germany. 
The institution offers a wide range of study programs across multiple disciplines. The study focused on the Department of Business Information Systems, including the bachelor’s programs Business Information Systems (BIS) and E-Government (EGOV). Both programs have a strong connection to computer science and information technology and include courses related to programming, information systems development, and digital transformation. Students in these programs frequently engage with digital tools and technologies as part of their academic work, making them a relevant population for investigating the use of GenAI tools.

In response to the emergence of GenAI technologies, the department has introduced initial institutional guidelines addressing the acceptable use of AI-supported tools in 2024 for academic work, particularly for written assignments and theses. However, as in many higher education institutions, these regulations are relatively recent and may not yet be fully understood or consistently communicated to students.

\subsection{Survey Design}
The questionnaire structure follows established principles of survey design as proposed by Fowler~\cite{Fowler.2014}. Following its initial development, the questionnaire was reviewed by the second and fourth co-authors of this paper.
Prior to the main data collection phase, the questionnaire was pilot-tested with a group of 20 students from the targeted study programs to improve clarity and comprehensibility of the survey instrument. Based on the feedback received, minor adjustments were made to the wording and structure of selected questions. The purpose of the reviews was to optimize the validity,  clarity, and relevance of the items. During this process, particular attention was paid to the wording of the items, the overall structure of the questionnaire, and the coverage of the underlying constructs. Based on the feedback received, slight revisions considering formulations were implemented to further enhance the quality of the questionnaire.
The questionnaire consisted of 15 questions and was structured into four thematic sections (the questionnaire is available here in~\cite{Bischof.2024}).


The first part of the questionnaire collected demographic and study-related information, including the participants’ study program, university affiliation, and current semester. The second part assessed students’ attitudes toward GenAI tools and their familiarity with these technologies. In addition, this part included a question measuring students’ awareness of institutional regulations governing the use of GenAI tools in their studies. The third part focused on students’ actual usage behavior of GenAI tools in their academic work. Participants were asked whether they use GenAI tools, which tools they use, and for which purposes they apply them within their studies. This part also captured students’ self-assessment regarding whether their use of GenAI tools complies with institutional regulations. The fourth and final part explored students’ perceived advantages and disadvantages of using GenAI tools in the study context and provided an optional opportunity for additional comments. While Fowler~\cite{Fowler.2014} recommends assessing behavior prior to attitudes, the present study prioritizes attitudes and awareness to avoid priming effects related to self-reported usage behavior. This ordering is justified by the research objective of capturing students’ perceptions independently from their reported usage patterns.

\subsection{Data Collection}
Data were collected through an anonymous online survey conducted during the winter semester 2024/2025. The survey targeted undergraduate students enrolled in the BIS and EGOV programs at the University of Applied Sciences and Arts Hannover, Germany.

The survey was conducted between $05.11.2024$ and $31.12.2024$, using the LimeSurvey tool. The questionnaire was distributed through multiple communication channels, including announcements during lectures, email invitations, and QR codes shared in relevant courses and online university platforms. Participation was voluntary and anonymous to ensure ethical compliance and reduce potential social desirability bias.\looseness=-1

In total, 151 valid responses were collected and included in the final dataset. The sample consisted of 93 students from the BIS program and 58 students from the EGOV program. Considering the total population of 465 BIS students and 115 EGOV students enrolled during the study period, the sample represents approximately $20\%$ of the BIS student population and around $50\%$ of the EGOV student population.

\subsection{Data Analysis}
The collected survey data were processed and analyzed using statistical software to ensure reliable data handling and evaluation. Data preparation and management were performed using the Python library Pandas. For the visualization of patterns and relationships within the dataset, the libraries Matplotlib and Seaborn were applied. The analysis focused on descriptive statistics and exploratory cross-tabulation of students’ awareness of institutional regulations and their perceived rule-compliant use of GenAI tools.

\textit{Sample Description:} In total, 151 students participated in the survey ($N=151$). All of the respondents were enrolled at the case university and belonged to one of the two study programs under investigation. The final sample included 93 students from the BIS program ($61.59\%$) and 58 students from the EGOV program ($38.41\%$). 
Table~\ref{tab:sample_distribution} provides an overview of the distribution of participating students across study terms and programs.

\begin{table}[h]
\centering
\begin{tabular}{c|c|c|c}
\toprule
\textbf{Semester} & \textbf{BIS Students} & \textbf{EGOV Students} & \textbf{Total} \\
\midrule
1 & 33 & 13 & 46 \\
2 & 16 & 2 & 18 \\
3 & 19 & 20 & 39 \\
4 & 5 & 0 & 5 \\
5 & 7 & 17 & 24 \\
6 & 6 & 0 & 6 \\
7 & 4 & 6 & 10 \\
8 & 3 & 0 & 3 \\
\midrule

\textbf{Total} & \textbf{93} & \textbf{58} & \textbf{N=151} \\
\bottomrule
\end{tabular}
\vspace{0.1cm}
\caption{Distribution of survey participants across semesters and study programs}
\label{tab:sample_distribution}
\end{table}

Overall, the sample distribution broadly reflects the general composition of the two programs. However, students enrolled in the fourth term were comparatively underrepresented. This imbalance can be explained by the fact that many students in this phase of their studies are completing mandatory internships outside the university during the survey period. 

With regard to the academic progression of participants, the majority of respondents were enrolled in the first to fourth semesters. Students in later semesters, particularly in the fifth and seventh terms, participated less frequently due to their involvement in off-campus practical training phases. Considering the total number of enrolled students in both programs, the achieved sample size represents a substantial share of the population. Specifically, the responses correspond to approximately $20\%$ of all BIS students (93 out of 465) and around $50\%$ of EGOV students (58 out of 115).

\textbf{Ethical Considerations}
To ensure compliance with ethical research standards, several measures were implemented during the study. Participation in the survey was entirely voluntary, and respondents could withdraw at any time without providing a reason. The survey was conducted anonymously, and no personally identifiable information was collected. Furthermore, the data collection process was approved by the university to ensure adherence to institutional ethical guidelines.

\subsection{Threats to Validity}
To provide transparency regarding potential sources of bias, we discuss the main limitations of this study based on the validity threat categories proposed by Runeson and Höst~\cite{Runeson.2009}.

\paragraph{Construct Validity}
In this study, the constructs of \textit{regulatory awareness} and \textit{perceived rule-compliant use} of GenAI tools were measured through self-reported survey responses. Respondents may have interpreted these concepts differently, particularly since institutional regulations regarding GenAI usage are still evolving and may not always be communicated consistently across courses. As a result, students’ understanding of what constitutes rule-compliant use may vary. Furthermore, because the study relies on self-reported behavior, responses may be affected by social desirability bias, even though the survey was conducted anonymously.

\paragraph{Internal Validity}
As the study is based on cross-sectional survey data, the results allow the identification of associations but do not permit causal conclusions. Several external factors may influence students’ awareness of regulations and their perceived rule-compliant use of GenAI tools. For example, students’ individual experience with digital tools or their exposure to discussions about GenAI within specific courses may affect both their awareness and their reported usage behavior. Additionally, the timing of the survey during the winter semester 2024/2025 may have influenced participation rates, particularly as students in later semesters were partially engaged in off-campus internships, which resulted in a lower representation of these groups.

\paragraph{External Validity}
The data were collected within a single higher education institution and focused on two computer science–related study programs. While these programs provide a relevant context for examining the use of GenAI tools in academic work, institutional policies, communication practices, and educational cultures may differ across universities and disciplines. Consequently, the results should be interpreted with caution when transferring them to other institutional contexts. Moreover, potential differences in students’ prior experience with digital technologies or their access to GenAI tools were not explicitly controlled for and may vary across institutions.

\paragraph{Conclusion Validity}
Our study primarily relies on descriptive statistics and exploratory analyses to examine relationships between regulatory awareness and perceived rule-compliant use. While these methods allow the identification of patterns within the dataset, they do not enable strong statistical generalizations beyond the investigated population. In addition, as participation in the survey was voluntary, self-selection effects may have occurred, potentially resulting in a higher participation rate among students who are more engaged with GenAI tools or more interested in the topic.

\section{Results}
\label{sec4:Results}
This section presents the empirical findings of our study. The results are structured according to the research questions. First, we analyze students’ awareness of institutional regulations governing GenAI tools. Second, we examine how students perceive the rule-compliance of their own GenAI usage. Third, we analyze the relationship between regulatory awareness and perceived rule-compliant behavior. Finally, we present a summary of the key findings before wie discuss our results in the next Section. 

\subsection{Awareness of Institutional GenAI Regulations}
Here, we answer RQ1: \textit{To what extent are students aware of institutional regulations governing the use of GenAI tools in higher education?}

Across the full sample of 151 respondents, only 39 students (25.83\%) reported that they were aware of such regulations. In contrast, 94 students (62.25\%) stated that they were not aware of the regulations, while 18 respondents (11.92\%) abstained from answering this question. Table~\ref{tab:regulation_awareness} depicts an overview of these results.

\begin{table}[h]
\centering
\begin{tabular}{p{2.8cm}ccc}
\toprule
\textbf{Awareness of Regulations} & \textbf{BIS} & \textbf{EGOV} & \textbf{Total} \\
\midrule
Known & 24 (25.81\%) & 15 (25.86\%) & 39 (25.83\%) \\
Unknown & 56 (60.22\%) & 38 (65.52\%) & 94 (62.25\%) \\
Abstention & 13 (13.98\%) & 5 (8.62\%) & 18 (11.92\%) \\
\midrule
\textbf{Total} & \textbf{93} & \textbf{58} & \textbf{N=151} \\
\bottomrule
\end{tabular}
\vspace{0.1cm}
\caption{Students’ awareness of institutional regulations governing the use of GenAI tools}
\label{tab:regulation_awareness}
\end{table}

A breakdown by study program shows only minor differences between the two groups. Among BIS students, 24 respondents (25.81\%) reported awareness of institutional regulations, 56 (60.22\%) reported no awareness, and 13 (13.98\%) abstained. Among EGOV students, 15 respondents (25.86\%) reported awareness, 38 (65.52\%) reported no awareness, and 5 (8.62\%) abstained. Overall, these results indicate that limited regulatory awareness is a shared pattern across both study programs rather than a program-specific phenomenon.

Thus, the dominant finding for RQ1 is that a clear majority of students in the investigated programs do not know the institutional rules governing GenAI use. This result is particularly notable given that the case institution had already introduced AI-related regulations and guidance for selected forms of academic work.

\subsection{Perceived Rule-Compliant Use of GenAI Tools}
Here, we answer RQ2:  \textit{How do students perceive the rule-compliance of their own use of GenAI tools in their academic
work? }

As shown in Table~\ref{tab:rule_compliance}, the question regarding rule-compliant use was answered only by respondents who reported using GenAI tools ($N=127$).
The results show a pronounced level of uncertainty. Only 22 students (17.32\%) indicated that their use of GenAI tools complies with the rules of their university. In contrast, 7 students (5.51\%) stated that their use was not rule-compliant. The largest group by far consisted of students who reported that they did not know whether their usage behavior was rule-compliant: 92 respondents (72.44\%). An additional 6 students (4.72\%) abstained.

\begin{table}[h]
\centering
\begin{tabular}{p{2.8cm}ccc}
\toprule
\textbf{Perceived Rule Compliance} & \textbf{BIS} & \textbf{EGOV} & \textbf{Total} \\
\midrule
Yes & 15 (18.75\%) & 7 (14.89\%) & 22 (17.32\%) \\
No & 3 (3.75\%) & 4 (8.51\%) & 7 (5.51\%) \\
I do not know & 57 (71.25\%) & 35 (74.47\%) & 92 (72.44\%) \\
Abstention & 5 (6.25\%) & 1 (2.13\%) & 6 (4.72\%) \\
\midrule
\textbf{Total} & \textbf{80} & \textbf{47} & \textbf{N=127} \\
\bottomrule
\end{tabular}
\vspace{0.1cm}
\caption{Students’ self-assessed rule-compliant use of GenAI tools}
\label{tab:rule_compliance}
\end{table}

The same pattern can be observed across both study programs. Among BIS students who used GenAI tools, 15 respondents (18.75\%) considered their use rule-compliant, 3 (3.75\%) considered it non-compliant, 57 (71.25\%) reported that they did not know, and 5 (6.25\%) abstained. Among EGOV students, 7 respondents (14.89\%) considered their use rule-compliant, 4 (8.51\%) considered it non-compliant, 35 (74.47\%) reported uncertainty, and 1 (2.13\%) abstained.

These findings indicate that uncertainty about rule-compliance is not a marginal phenomenon. Rather, it represents the dominant response pattern among students who actively use GenAI tools. In substantive terms, the results suggest that many students continue to use GenAI in their studies even though they are unable to determine whether their own usage aligns with institutional expectations.

\subsection{Relationship Between Awareness and Compliance}
Here, we answer RQ3: 
\textit{What relationship exists between students’ awareness of institutional GenAI regulations and their perceived rule-compliant use of these tools?}

\begin{figure}[thb]
\centering
	\includegraphics[scale=0.29]{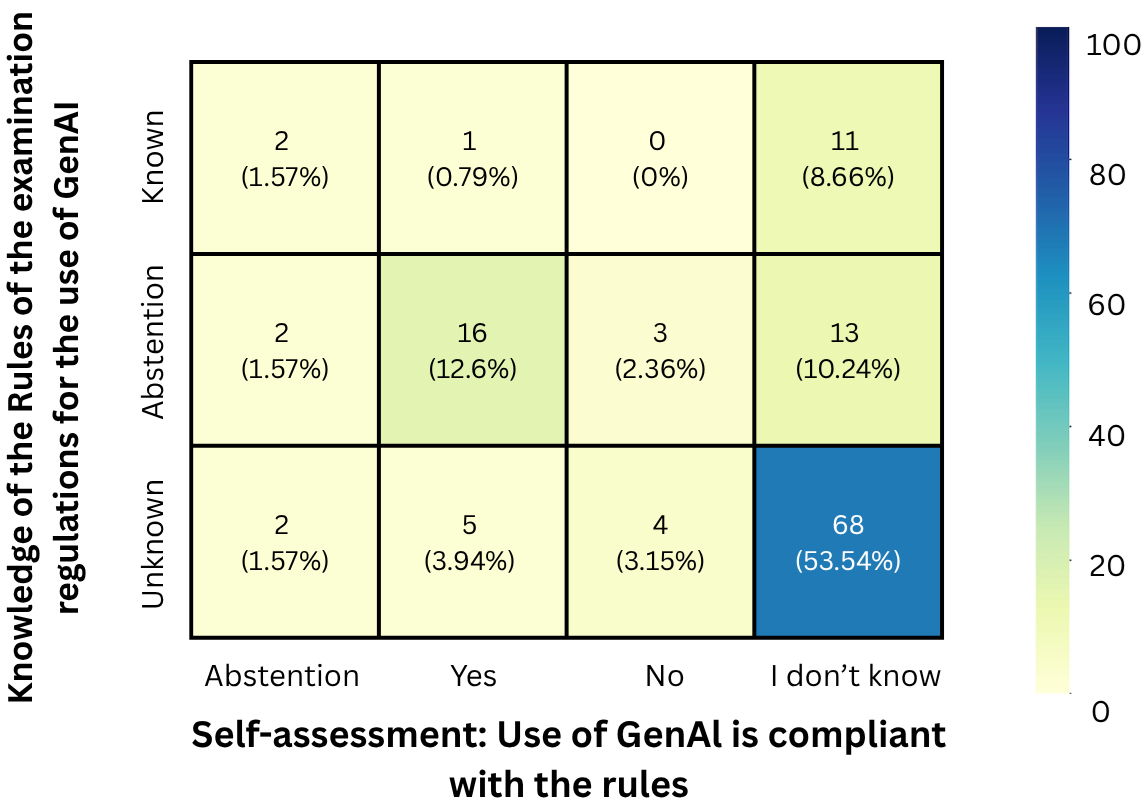}
\caption{Relationship between awareness of institutional GenAI regulations and perceived rule-compliant use}
\label{fig:awareness}
\end{figure}

To investigate this relationship, we analyzed the cross-tabulated distribution of the two variables among students who reported using GenAI tools. Figure~\ref{fig:awareness} visualizes this relationship using a heatmap, where the vertical axis represents students’ awareness of institutional regulations and the horizontal axis represents their self-assessment regarding whether their use of GenAI tools complies with these rules.

The heatmap reveals a clear concentration of responses in the category \textit{I do not know} with regard to rule-compliant use. Among the students who reported using GenAI tools, 92 out of 127 respondents (72.44\%) indicated that they were unable to assess whether their usage complies with institutional regulations. By contrast, only 22 students (17.32\%) stated that their use is rule-compliant, while 7 students (5.51\%) reported that their use does not comply with institutional rules.

A closer examination of the heatmap shows that students who reported being unfamiliar with institutional regulations are strongly concentrated in the \textit{I do not know} category. This indicates that a lack of awareness of institutional rules is closely associated with uncertainty regarding acceptable use. However, the results also show that even among students who reported awareness of institutional regulations, certainty about rule-compliant behavior remains limited.

Overall, the findings point to a clear descriptive pattern: students with limited awareness of institutional regulations are frequently unable to assess whether their use of GenAI tools is rule-compliant. However, the results also show that awareness does not automatically lead to actionable clarity. The heatmap indicates that uncertainty about acceptable use persists even among students who reported being familiar with the institutional rules. This suggests that students may know that regulations exist, while still lacking a clear understanding of how these rules apply to concrete practices such as writing, programming support, or the preparation of assignments. The central empirical pattern therefore points to a broader gap between institutional regulation and students’ ability to interpret and apply these rules in their everyday academic use of GenAI tools.

\subsection{Summary of Key Findings}
This section summarizes the main empirical findings of the study.

\vspace{0.2cm}
\noindent
\fbox{
\begin{minipage}{0.95\linewidth}
\textbf{Key Findings.}
The results indicate that awareness of institutional regulations regarding GenAI tools is limited among students. At the same time, most students who use GenAI are uncertain whether their usage complies with these regulations. Overall, the findings highlight a gap between the existence of institutional rules and students’ ability to confidently assess rule-compliant use.
\end{minipage}
}
\vspace{0.2cm}

First, the results show that awareness of institutional regulations governing the use of GenAI tools is generally low among students. Only 39 out of 151 respondents (25.83\%) reported that they were aware of such regulations, while the majority of students (62.25\%) indicated that they were not familiar with these rules.

Second, among students who reported using GenAI tools in their studies, a large proportion expressed uncertainty regarding the rule-compliance of their own usage. Of the 127 respondents in this subset, 92 students (72.44\%) stated that they did not know whether their use of GenAI tools complies with institutional regulations.

Third, the cross-analysis of regulatory awareness and perceived rule-compliant use suggests that awareness of institutional regulations does not necessarily translate into clarity regarding acceptable use.

\section{Discussion}
\label{sec5:Discussion}
The results of this study reveal several important insights into how students perceive and interpret institutional regulations governing the use of GenAI tools in higher education. In particular, the findings highlight a notable gap between the existence of institutional regulations and students’ awareness and understanding of these rules. This observation extends prior work on students’ AI awareness, ethical reflection, and regulatory perception in higher education (\textit{e.g.}, \cite{Ghotbi.2021,Dergunova.2022,Simon.2024,Wu.2025,Tan.2025}).

Our results indicate that awareness of institutional regulations regarding GenAI tools is limited among students. Only 39 out of 151 respondents (25.83\%) reported that they were aware of such regulations, while 94 students (62.25\%) stated that they were not familiar with these rules. This finding suggests that the mere existence of institutional policies does not necessarily ensure that students are aware of them. This aligns with prior work showing that regulation and governance of AI in higher education are becoming increasingly important, while their practical interpretation remains challenging (\textit{e.g.}, \cite{Miah.2025,Stephano_2025,Tan.2025}).

Furthermore, we identified a high degree of uncertainty among students regarding whether their own use of GenAI tools complies with institutional rules. Among the 127 respondents who reported using GenAI tools in their studies, 92 students (72.44\%) indicated that they did not know whether their usage behavior was rule-compliant. Only 22 students (17.32\%) reported that their use complies with institutional regulations. 
This widespread uncertainty suggests that existing guidelines may be perceived as unclear, difficult to interpret, or insufficiently contextualized for specific academic tasks such as assignments, programming exercises, or written reports. In this sense, our findings complement prior studies showing that students’ engagement with AI is shaped not only by adoption, but also by awareness, reflection, and ethical considerations (\textit{e.g.}, \cite{Ghotbi.2021,Ferhataj.2025,Liu.2025,Wu.2025}).
This issue is particularly relevant because the respondents in our study are bachelor students who are still learning what academic integrity means in practice.
In this phase of academic socialization, students are expected to develop an understanding of responsible authorship, transparency in the use of tools, and the boundaries of acceptable academic support. If they do not know the applicable rules for GenAI use, or if these rules remain unclear to them, this does not only affect isolated usage decisions. It may also shape how they develop academic judgment and how they understand responsible scholarly work more broadly.

The cross-analysis of regulatory awareness and perceived rule-compliant use indicates that awareness of institutional regulations alone does not necessarily translate into clarity regarding acceptable behavior. 
This finding refines prior work showing that awareness influences students’ perceptions of GenAI, but also suggests that awareness alone is not sufficient for confident rule interpretation in practice (\textit{e.g.}, \cite{Simon.2024,Wu.2025,Tan.2025}).
Even among students who reported awareness of institutional rules, uncertainty regarding rule-compliant use remained common. This finding is particularly noteworthy in a broader European higher education context, where institutional guidance on GenAI use is embedded in a comparatively dense regulatory environment shaped by, among others, the GDPR and the EU AI Act. 
The persistence of uncertainty under these conditions suggests that the central challenge is not merely the existence or awareness of rules, but students’ ability to interpret and apply them in concrete academic situations. At the same time, this contextual embeddedness also limits straightforward generalization: in higher education settings outside Europe, or in institutional contexts with less prominent external regulation, different effects may emerge, either because fewer formal constraints are communicated or because acceptable use is governed more informally.

In summary, the results point to a broader disconnect between institutional regulatory frameworks and students’ everyday learning practices. GenAI tools are already widely embedded in students’ academic workflows, yet institutional governance mechanisms appear to lag behind these evolving practices. As a consequence, students may continue to use these tools despite uncertainty about whether their behavior aligns with institutional expectations.

From a practical perspective, the findings suggest that higher education institutions should complement regulatory approaches with clearer communication strategies and pedagogical guidance. Instead of relying solely on formal regulations, institutions may benefit from integrating discussions about acceptable AI use directly into courses, assignments, and assessment guidelines. For example, instructors could provide explicit examples of permitted and non-permitted uses of GenAI tools in specific learning contexts.

Also, the high level of uncertainty observed in the survey indicates a need for greater transparency regarding how GenAI tools may be used in academic work. Institutions should consider developing accessible guidelines, training materials, or workshops that help students better understand how to use these technologies responsibly and in accordance with academic integrity principles.

Overall, the findings of our study suggest that effective governance of GenAI in higher education may require a combination of regulatory frameworks, clear communication, and pedagogical integration. Rather than focusing exclusively on restricting AI use, institutions may benefit from supporting students in developing informed and responsible practices for using GenAI tools in their studies.

\section{Conclusion \& Future Work}
\label{sec6:Conclusion}
In this study, we investigated the relationship between students’ awareness of institutional regulations governing the use of GenAI tools and their perceived rule-compliant use in higher education. Based on a quantitative survey of 151 undergraduate students enrolled in Business Information Systems and E-Government programs, the study provides empirical insights into how students perceive and interpret institutional GenAI regulations.

The results indicate that awareness of institutional regulations governing GenAI use is limited among students. Only 39 out of 151 respondents (25.83\%) reported being aware of such regulations, while 94 students (62.25\%) stated that they were not familiar with them. At the same time, GenAI tools are already widely used in students’ academic work. Among the respondents who reported using GenAI tools, 92 out of 127 students (72.44\%) indicated that they were uncertain whether their usage complies with institutional regulations. Furthermore, the analysis of the relationship between regulatory awareness and perceived rule-compliant behavior indicates that awareness of institutional regulations alone does not necessarily translate into clarity regarding acceptable AI use. Even students who reported familiarity with institutional rules often remained uncertain about how these regulations apply in practice.

In conclusion, this study contributes empirical evidence to the ongoing discussion about the governance of GenAI in higher education. The findings emphasize that establishing regulations alone may not be sufficient to guide student behavior. Instead, institutions may need to complement regulatory approaches with clearer communication strategies and practical guidance on how GenAI tools can be used responsibly in academic work.

Future research could extend this work in several directions. First, similar studies could be conducted at other universities and across different disciplines to examine whether the observed patterns are consistent across institutional and educational contexts. Second, longitudinal studies may investigate how students’ awareness and perceptions evolve as institutional AI policies mature and become more integrated into teaching practices. We also call other researchers to explore how different approaches to communicating and integrating AI regulations into courses influence students’ understanding of responsible GenAI use.

\section*{Acknowledgment}
We would like to thank the students who participated in our study. 

\bibliographystyle{IEEEtran}
\bibliography{references}

@article{von_garrel_kunstliche_2023,
	title = {Künstliche Intelligenz im Studium Eine quantitative Befragung von Studierenden zur Nutzung von {ChatGPT} \&amp; Co.},
	rights = {Creative Commons - {CC} {BY} - Namensnennung 4.0 International},
	url = {https://opus4.kobv.de/opus4-h-da/395},
	doi = {10.48444/H_DOCS-PUB-395},
	author = {von Garrel, Jörg and Mayer, Jana and Mühlfeld, Markus},
	urldate = {2024-12-28},
	year = {2023},
	langid = {german}
}

@article{gottschling_nutzung_2024,
	title = {Nutzung von {KI}-Tools durch Studierende},
	issn = {{ISSN}: 2199-8825},
	doi = {DOI: 10.3278/HSL2411W},
	number = {11},
	journaltitle = {die hochschullehre},
	author = {Gottschling, Steffen and Seidl, Tobias and Vonhof, Cornelia},
	year = {2024},
	langid = {german},
}

@INPROCEEDINGS{neumann_we_2023,
  author={Neumann, Michael and Rauschenberger, Maria and Schön, Eva-Maria},
  booktitle={Proc. of the 5th Intnl. Workshop on Software Engineering Education for the Next Generation}, 
  title={“We Need To Talk About ChatGPT”: The Future of AI and Higher Education}, 
  year={2023},
  volume={},
  number={},
  pages={29-32},

  doi={10.1109/SEENG59157.2023.00010}}

@inproceedings{Ruedian.2025,
author = {R\"{u}dian, Sylvio and Podelo, Julia and Ku\v{z}\'{\i}lek, Jakub and Pinkwart, Niels},
title = {Feedback on Feedback: Student’s Perceptions for Feedback from Teachers and Few-Shot LLMs},
year = {2025},
isbn = {9798400707018},
publisher = {Association for Computing Machinery},
address = {New York, NY, USA},
doi = {10.1145/3706468.3706479},
booktitle = {Proceedings of the 15th International Learning Analytics and Knowledge Conference},
pages = {82–92},
numpages = {11},
series = {LAK '25}
}

@ARTICLE{Shailendra.2024,
  author={Shailendra, Samar and Kadel, Rajan and Sharma, Aakanksha},
  journal={IEEE Transactions on Education}, 
  title={Framework for Adoption of Generative Artificial Intelligence (GenAI) in Education}, 
  year={2024},
  volume={67},
  number={5},
  pages={777-785},
  doi={10.1109/TE.2024.3432101}}

@INPROCEEDINGS{Speth.2023,
  author={Speth, Sandro and Meißner, Niklas and Becker, Steffen},
  booktitle={Proc. of the 35th Intnl. Conf. on Software Engineering Education and Training}, 
  title={Investigating the Use of AI-Generated Exercises for Beginner and Intermediate Programming Courses: A ChatGPT Case Study}, 
  year={2023},
  volume={},
  number={},
  pages={142-146},
  doi={10.1109/CSEET58097.2023.00030 }}

@inproceedings{Meissner.2024,
author = {Mei\ss{}ner, Niklas and Koch, Nadine and Speth, Sandro and Breitenb\"{u}cher, Uwe and Becker, Steffen},
title = {Unveiling Hurdles in Software Engineering Education: The Role of Learning Management Systems},
year = {2024},
isbn = {9798400704987},
publisher = {Association for Computing Machinery},
address = {New York, NY, USA},
doi = {10.1145/3639474.3640060},
booktitle = {Proc. of the 46th Intnl. Conf. on Software Engineering: Software Engineering Education and Training},
pages = {242–252},
numpages = {11},
location = {Lisbon, Portugal},
series = {ICSE-SEET '24}
}

@inproceedings{Daun.2023,
author = {Daun, Marian and Brings, Jennifer},
title = {How ChatGPT Will Change Software Engineering Education},
year = {2023},
isbn = {9798400701382  },
publisher = {Association for Computing Machinery},
address = {New York, NY, USA},
doi = {10.1145/3587102.3588815   },
booktitle = {Proceedings of the 2023 Conference on Innovation and Technology in Computer Science Education V. 1},
pages = {110–116},
numpages = {7},
location = {Turku, Finland},
series = {ITiCSE 2023}
}

@ARTICLE{Schoen.2023, 
AUTHOR={Sch{\"o}n, Eva-Maria  and Neumann, Michael  and Hofmann-St{\"o}lting, Christina  and Baeza-Yates, Ricardo  and Rauschenberger, Maria },
TITLE={How are AI assistants changing higher education?},
JOURNAL={Frontiers in Computer Science},
VOLUME={5},
YEAR={2023},
DOI={10.3389/fcomp.2023.1208550 },
ISSN={2624-9898}
}

@INPROCEEDINGS{Lauren.2023,
  author={Lauren, Paula and Watta, Paul},
  booktitle={2023 IEEE Frontiers in Education Conference (FIE)}, 
  title={Work-in-Progress: Integrating Generative AI with Evidence-based Learning Strategies in Computer Science and Engineering Education}, 
  year={2023},
  volume={},
  number={},
  pages={1-5},
  doi={10.1109/FIE58773.2023.10342970}}

@INPROCEEDINGS{Maher.2023,
  author={Maher, Mary Lou and Tadimalla, Sri Yash and Dhamani, Dhruv},
  booktitle={Proc. of the Frontiers in Education Conference}, 
  title={An Exploratory Study on the Impact of AI tools on the Student Experience in Programming Courses: an Intersectional Analysis Approach}, 
  year={2023},
  volume={},
  number={},
  pages={1-5},
  doi={10.1109/FIE58773.2023.10343037}}

@INPROCEEDINGS{Groepler.2025,
author={Gr{\"o}pler, Robin and Klepke, Steffen and Johns, Jack and Dreschinski, Andreas and Schmid, Klaus and Dornauer, Benedikt and T{\"u}z{\"u}n, Eray and Noppen, Joost and Mousavi, Mohammad Reza and Tang, Yongjian and Viehmann, Johannes and Aslang{\"u}l, Selin {\c S}irin and Lee, Beum Seuk and Ziolkowski, Adam and Zie, Eric},
  booktitle={Proceedings of the 2nd IEEE/ACM International Conference on AI-powered Software (AIware)}, 
  title={The Future of Generative AI in Software Engineering: A Vision From Industry and Academia in the European Genius Project}, 
  year={2025},
  volume={},
  number={},
  pages={170-181},
  doi={10.1109/AIware69974.2025.00026}}

@article{Runeson.2009,
 author = {Runeson, P. and H{\"o}st, M.},
 year = {2009},
 title = {Guidelines for conducting and reporting case study research in software engineering},
 pages = {131--164},
 volume = {14},
 number = {2},
 issn = {1382-3256},
 journal = {Empirical Software Engineering},
 doi = {\url{10.1007/s10664-008-9102-8}}
}

@article{Southworth.2023,
title = {Developing a model for AI Across the curriculum: Transforming the higher education landscape via innovation in AI literacy},
journal = {Computers and Education: Artificial Intelligence},
volume = {4},
pages = {100127},
year = {2023},
issn = {2666-920X},
doi = {https://doi.org/10.1016/j.ca          eai.2023.100127}         ,
author = {Jane Southworth and Kati Migliaccio and Joe Glover and Ja’Net Glover and David Reed and Christopher McCarty and Joel Brendemuhl and Aaron Thomas}
}

@article{Su.2023,
 author = {J. Su and W. Yang},
 year = {2023},
 title = {Unlocking the power of ChatGPT: A framework for applying generative AI in education},
 pages = {355–366},
 volume = {6},
 number = {3},
 journal = {ECNU Rev. Educ.}
}

@misc{Bischof.2024,
author = {Lasse Bischof and Maria Rauschenberger and Eva-Maria Schön and Michael Neumann},
year = {2024},
url = {https://doi.org/10.5281/zenodo.15017474                                                    },
title = {Questionnaire GenAI usage by students}
}

@conference{Bischof.2025,
author={Lasse Bischof and Eva-Maria Schön and Maria Rauschenberger and Michael Neumann},
title={“We Need to Analyze Students GenAI Use”: Towards an AI Adoption Framework for Higher Education},
booktitle={Proceedings of the 21st International Conference on Web Information Systems and Technologies},
year={2025},
pages={429-438},
publisher={SciTePress},
organization={INSTICC},
doi={10.5220/0013819000003985},
isbn={978-989-758-772-6},
}

@article{Aldossary.2024,
 author = {Aldossary, A. S. and Aljindi, A. A. and Alamri, J. M.},
 year = {2015},
 title = {The role of generative AI in education: Perceptions of Saudi students},
 pages = {},
 volume = {16},
 number = {4},
 journal = {Contemporary Educational Technology},
 doi = {\url{10.30935/cedtech/15496}}
}

@article{Mah.2024,
  author       = {Mah, D. K. and Groß, N.},
  title        = {Artificial intelligence in higher education: exploring faculty use, self-efficacy, distinct profiles, and professional development needs},
  journal      = {International Journal of Educational Technology in Higher Education},
  year         = {2024},
  volume       = {21},
  pages        = {58},
  doi          = {10.1186/s41239-024-00490-1}
}

@article{Wang.2025,
  author       = {Wang, C.},
  title        = {Exploring Students’ Generative AI-Assisted Writing Processes: Perceptions and Experiences from Native and Nonnative English Speakers},
  journal      = {Technology, Knowledge and Learning},
  year         = {2025},
  volume       = {30},
  pages        = {1825--1846},
  doi          = {10.1007/s10758-024-09744-3},
  url          = {https://doi.org/10.1007/s10758-024-09744-3}
}

@article{Rizki.2025,
  author       = {Rizki, I.~A. and Daoud, R.},
  title        = {Generative Artificial Intelligence in Higher Education: Review of Institutional Policies and Practices across New Zealand},
  journal      = {New Zealand Journal of Educational Studies},
  year         = {2025},
  doi          = {10.1007/s40841-025-00417-y},
  url          = {https://doi.org/10.1007/s40841-025-00417-y}
}

@article{Wiese.2025,
title = {AI ethics education: A systematic literature review},
journal = {Computers and Education: Artificial Intelligence},
volume = {8},
pages = {100405},
year = {2025},
issn = {2666-920X},
doi = {https://doi.org/10.1016/j.caeai.2025.100405},
url = {https://www.sciencedirect.com/science/article/pii/S2666920X25000451},
author = {Lucas J. Wiese and Indira Patil and Daniel S. Schiff and Alejandra J. Magana},
}

@ARTICLE{Kitchenham.2022,
  author={Kitchenham, B.A. and Pfleeger, S.L. and Pickard, L.M. and Jones, P.W. and Hoaglin, D.C. and El Emam, K. and Rosenberg, J.},
  journal={IEEE Transactions on Software Engineering}, 
  title={Preliminary guidelines for empirical research in software engineering}, 
  year={2002},
  volume={28},
  number={8},
  pages={721-734},
  doi={10.1109/TSE.2002.1027796}}

@article{Miah.2025, 
title={Regulating Artificial Intelligence in Education: Analyzing Legal and Ethical Frameworks for the Deployment of AI and Machine Learning Models in U.S. Educational Institutions}, volume={7},
number={11}, 
journal={Journal of Computer Science and Technology Studies}, 
author={Miah, Mohammed Nazmul Islam and Uddin, Md Joshim and Ahmed, Md Wasim}, 
year={2025}, 
month={Nov.}, 
pages={387–404} }

@article{Ghotbi.2021, 
title={Moral Awareness of College Students Regarding Artificial Intelligence}, 
volume={13},
doi={10.1007/s41649-021-00182-2}, 
journal={Asian Bioethics Review}, 
author={Ghotbi, N. and Ho, M.T.}, 
year={2021}, 
pages={421-433} }

@article{Dergunova.2022,
	author = { Yelena Dergunova and Rakhila Aubakirova and Botagoz Yelmuratova and Tulekova Gulmira and Pigovayeva Yuzikovna and Samal Antikeyeva },
	title = { Artificial Intelligence Awareness Levels of Students },
	journal = { International Journal of Emerging Technologies in Learning (iJET) },
	volume = { 17 },
	number = { 18 },
	year = { 2022 },
	month = { September },
	pages = { 26--37 },
	address = { Kassel, Germany },
	publisher = { International Journal of Emerging Technology in Learning },
	issn = { 1863-0383 },
	url = { https://www.learntechlib.org/p/223076 }
}

@article{Ferhataj.2025,
    author = {Ferhataj, Anxhela and Memaj, Fatmir and Sahatcija, Roland and Ora, Ariel and Koka, Enkelejda},
    title = {Ethical concerns in AI development: analyzing students’ perspectives on robotics and society},
    journal = {Journal of Information, Communication and Ethics in Society},
    volume = {23},
    number = {2},
    pages = {165-187},
    year = {2025},
    month = {01},
    issn = {1477-996X},
    doi = {10.1108/JICES-08-2024-0111},
    url = {https://doi.org/10.1108/JICES-08-2024-0111}
}

@article{Liu.2025,
    author = {Liu, X. and Xiao, Y. and Li, D.},
    title = {Assessing strategic use of artificial intelligence in self-regulated learning: Instrument development and evidence from Chinese university students},
    journal = {Int J Educ Technol High Educ},
    volume = {22},
    number = {69},
    pages = {165-187},
    year = {2025},
    doi = {10.1186/s41239-025-00567-5}
}

@Inbook{Simon.2024,
author="Simon, Simi
and Suresh, Samyuktha Paliathuparambil
and Nithyananda, Santhosh
and Unnikrishnan, Ramesh Kumar
and Suresh, Sangama Paliyathparambil
and Manayath, Dhanya",
editor="Mansour, Nadia
and Bujosa Vadell, Lorenzo M.",
title="Exploring the Role of Generative AI Awareness in Shaping Student Perceptions",
bookTitle="Finance and Law in the Metaverse World: Regulation and Financial Innovation in the Virtual World",
year="2024",
publisher="Springer Nature Switzerland",
address="Cham",
pages="519--531",
isbn="978-3-031-67547-8",
doi="10.1007/978-3-031-67547-8_44"
}

@article{Wu.2025,
    author = {Wu, H. and Li, D. and Mo, X.},
    title = {Understanding GAI risk awareness among higher vocational education students: An AI literacy perspective},
    journal = {Educ Inf Technol},
    volume = {30},
    pages = {14273–14304},
    year = {2025},
    doi = {10.1007/s10639-024-13312-8}
}

@article{Tan.2025,
author = {Tan, Michelle Xin Yi and Qu, Yao and Wang, Jue},
title = {Student Perceptions of Generative Artificial Intelligence Regulations: A Mixed-Methods Study of Higher Education in Singapore},
journal = {Higher Education Quarterly},
volume = {79},
number = {3},
pages = {e70038},
doi = {https://doi.org/10.1111/hequ.70038},
year = {2025}
}

@article{Stephano_2025, 
title={Evaluating the Extent of Copyright Law Awareness Among Lecturers in Regulating AI Use by Students in Higher Education }, 
volume={16}, 
number={2}, 
journal={NG Journal of Social Development}, 
author={Stephano , Godwin}, 
year={2025}, month={Apr.}, 
pages={63–74},
doi={10.4314/ngjsd.v16i2.5}
}

@article{Wang.2024,
title = {Artificial intelligence in education: A systematic literature review},
journal = {Expert Systems with Applications},
volume = {252},
pages = {124167},
year = {2024},
issn = {0957-4174},
doi = {https://doi.org/10.1016/j.eswa.2024.124167},
author = {Shan Wang and Fang Wang and Zhen Zhu and Jingxuan Wang and Tam Tran and Zhao Du}
}

@article{Chiu.2023,
title = {Systematic literature review on opportunities, challenges, and future research recommendations of artificial intelligence in education},
journal = {Computers and Education: Artificial Intelligence},
volume = {4},
pages = {100118},
year = {2023},
issn = {2666-920X},
doi = {https://doi.org/10.1016/j.caeai.2022.100118},
author = {Thomas K.F. Chiu and Qi Xia and Xinyan Zhou and Ching Sing Chai and Miaoting Cheng}
}

@book{Fowler.2014,
  author    = {Floyd J. Fowler Jr.},
  title     = {Survey Research Methods},
  edition   = {5},
  year      = {2014},
  publisher = {SAGE Publications},
  isbn      = {9781452259000}
}

@inproceedings{Mendonca.2025,
  author    = {Paula Mendon{\c{c}}a and Jos{\'e} Reginaldo Hughes Carvalho and A. Oran},
  title     = {Using Prompt Engineering to Enhance a Project-Based Learning Course on Project Management},
  booktitle = {Proceedings of the 55th IEEE Frontiers in Education Conference},
  year      = {2025},
  month     = nov,
  pages     = {1--9},
  doi       = {10.1109/FIE63693.2025.11328531}
}

@inproceedings{Tarun.2025,
  author    = {Bhavishya Tarun and Haoze Du and Dinesh Kannan and Edward F. Gehringer},
  title     = {Human-in-the-Loop Systems for Adaptive Learning Using Generative AI},
  booktitle = {Proceedings of the 55th IEEE Frontiers in Education Conference},
  year      = {2025},
  month     = nov,
  pages     = {1--7},
  doi       = {10.1109/FIE63693.2025.11328658}
}

@inproceedings{DeFrancisis.2025,
  author    = {David A. DeFrancisis and David Pabst and Lee A. Dosse and Jacklyn F. Wyszynski and Matthew M. Barry},
  title     = {Exploring Undergraduate Students' Utilization and Perceptions of Generative AI in Engineering: Insights from an Introductory Statics and Mechanics of Materials Course},
  booktitle = {Proceedings of the 55th IEEE Frontiers in Education Conference},
  year      = {2025},
  month     = nov,
  doi       = {10.1109/FIE63693.2025.11328640}
}

@inproceedings{Li.2025,
  author    = {Zepei Li and Sotiria Koloutsou-Vakakis and Tomasz Kozlowski and Volodymyr Kindratenko and Abdussalam Alawini},
  title     = {How Students Use Generative AI: Insights from Conversation Log Analysis},
  booktitle = {Proceedings of the 55th IEEE Frontiers in Education Conference},
  year      = {2025},
  month     = nov,
  pages     = {1--7},
  doi       = {10.1109/FIE63693.2025.11328645}
}

\end{document}